\documentclass[3p,times,procedia]{elsarticle}
\usepackage{ecrc}

\volume{00}

\firstpage{1}

\journalname{Procedia IUTAM}

\runauth{ Tihomir Hristov}

\jid{piutam}

\usepackage{amssymb}

\usepackage[figuresright]{rotating}

\begin{document}
\begin{frontmatter}



\dochead{IUTAM Symposium Wind Waves, 4-8 September 2017, London, UK}%

\title{Mechanistic, empirical and numerical perspectives \\
		on wind-waves interaction}


\author{Tihomir Hristov} 

\address[a]{Department of Mechanical Engineering, Johns Hopkins University,   
    3400 N.\ Charles Str., Baltimore, MD 21218, USA  }

\begin{abstract}    

  A mechanistic theory of wind-wave interaction must rely on
  verifiable assumptions and offer reproducible observable
  predictions.  For decades, the limited mechanistic grasp on the
  problem has motivated RANS and LES modeling and has driven a vast
  empirical effort to describe the interaction in terms of
  wave-induced modifications of standard statistical characteristics
  of the wind, such as wind profile, kinetic energy balance or
  exchange coefficients.  Because the mechanistic, empirical and
  numerical approaches are all concerned with the same phenomenon
  occurring in the same media, consistency here requires that the
  assumptions on which the approaches rest and the predictions they
  generate are compatible with each other and supported by
  measurements. Recent findings from theoretical analysis and field
  experiments advanced the understanding of the statistical and
  dynamic patterns of the wave-coherent flow, which is at the core of
  the mechanistic description of the wind-wave exchange.  The progress
  prompts reexamining of earlier concepts, efforts and findings to
  evaluate their suitability, validity and usefulness.  For the
  purpose, this survey traces the development of ideas, methods and
  results in the study of the wind wave generation.

\end{abstract}

\end{frontmatter}

\correspondingauthor[*]{Tihomir Hristov}
\email{Tihomir.Hristov@jhu.edu, Hristov.Tihomir@gmail.com}



\section{Introduction}

'What causes the waves in the ocean?' is a question that seeks to
elucidate an ubiquitous phenomenon and in that sense it stands next to
questions such as 'What makes the stars shine?' or 'What causes the
lightning?'.  In his thesis advised by Richard Feynman, "The growth of
water waves due to the action of the wind", \citet{HibbsWaves}
wondered about {\em "Why should such a problem still exist, with all
  the abilities of modern physics and the accomplishments of modern
  aerodynamics?".}  Indeed, initiated by \citet{kelvin}, the efforts
to reveal the physics behind the wind-generated ocean waves are still
ongoing, as evidenced by this meeting. In his influential review on
the subject \citet[p.\ 217]{UrsellReview1956} pointed out the key
obstacle to understanding the process that is the disconnect between
theory and experiment, a disconnect that began to close only recently.
Since the times of \citet{UrsellReview1956}, however, the lines of
inquiry on wind-wave interaction have been broadened beyond theory and
experiment to include empirical and numerical approaches which, like
theory and experiment, have also developed separately and any possible
connection between them has stayed unexplored. As all these approaches
are concerned with the same phenomenon in the same environment, the
results and predictions from different approaches should be consistent
with each other.  Whether a theory should be about a mechanism or
unity, as \citet[p.\ 177]{Poincare_ScienceHypothesis} pondered, this
survey pursues both.  Because a mechanistic theory on wind-wave
interaction should be capable to provide details and physical insight,
a summary of a theory is used below to gain an unified mechanistic
frame of reference for empirical and numerical approaches and results.  For
the purpose, here we discuss the corollaries of some recent findings
on the problem of waves generation by wind, the analytic and
experimental details of which are available in \cite{hristovNature},
\cite{HristovJPO2014} and \cite{Hristov_WileyEncyclopedia}.  To put
these findings in perspective and to answer the question raised by Hibbs'
\cite{HibbsWaves}, we outline the milestones in the early
evolution of ideas as they may, justifiably or not, still guide the
thinking on the subject.

\section{Milestones in the early wind-wave interaction studies}

Like \citet{kelvin}, \citet{jeffreys24} studied the process of wind
waves generation assuming an air flow velocity constant with the
distance from the interface. At the time the specific functional form of 
the wind velocity profile was still unknown and its role 
in the generation had not  been studied and understood.
Starting from such unrealistic air flow \citet{jeffreys24} arrived to
unrealistic predictions and incorrectly traced them back to the
assumption of irrotational waves.  Abandoning that assumption while
keeping the unrealistic wind profile, \citet{jeffreys24} proceeded to
hypothesize and prescribe, rather than to derive or otherwise deduce,
the physical nature of the interaction.  Forty years later \citet[p.
S48]{StewartRW_PhysFluids1967} found \citet{jeffreys24} solution of
the wind wave generation problem lacking.

The review of \citet{UrsellReview1956} summarized the ideas available 
at the time, stated the unsatisfactory understanding of the problem,  
and inspired the works of \citet{phillips57} and \citet{miles}.
\citet{phillips57} built on the work by \citet{EckartWaveGeneration} 
  and considered a wave
generation due to the turbulent air pressure fluctuations as in an
oscillator under the action of a random force.  Paradoxically,
\citet[p.\ 417]{phillips57} both neglected the wave-coherent motion in the air 
and found it to be most effective for the generation process:
{\em
  "Correlations between air and water motions are neglected ...  It is
  found that waves develop most rapidly by means of a resonance
  mechanism which occurs when a component of the surface pressure
  distribution moves at the same speed as the free surface wave with
  the same wavenumber."} For the resonant
wave growth to occur, \citet[p.\ 422]{phillips57} required that {\em
  "the pressure distribution contains components whose wave-numbers
  and frequencies coincide with possible modes of free surface
  waves"}.  
Yet, the dispersion of the turbulent motion in a wind with advection
velocity $U$ is $U = \omega/k$, also known as Taylor's hypothesis, and
the dispersion of surface waves with a phase speed $c$ is $c =
g/\omega$, $g$ being the acceleration of gravity.  Because for aligned
wind and waves same frequency and wave number correspond to different
advection velocity $U$ and phase speed $c$, these dispersions are
incompatible for resonance. For a wave mode $\omega(k)$ a resonance
may still occur if it propagates obliquely at some angle $\pm \theta$
with respect to the wind, thus predicting a wave field with a bimodal
spectral distribution. The observations of \citet[p.\
93]{JHSimpson1969} detected no bimodal spectrum in the range of the
locally generated waves.  \citet{Longuet-Higgins62} reported the
turbulent air pressure fluctuations to be too small to ensure wave
growth rates near the observed, but offered no interpretation. The
turbulent pressure fluctuations decay with scale as $k^{-7/3}$
\citep{ObukhovPressure,BatchelorPressure}, that is faster than the
decay rate with scale $k^{-5/3}$ in velocity.  As turbulent motion
undergoes the energy cascade from integral scales down to the scales
of the initiating sea surface ripples, only small magnitude pressure
fluctuations are retained. Consequently, the turbulent pressure
component on which the random force mechanism relies, is virtually
absent, explaining the findings of \citet{Longuet-Higgins62} and
causing the reported lack of experimental support for the random
forcing mechanism of wave generation \citep{JHSimpson1969}.

The idea that the generation of wind-driven waves may be viewed as a
shear flow instability has its roots in the work of
\citet{tollmien_productionofturbulence}, published as an obscure
technical memorandum for NACA and thus rarely referenced.  Its results
became available to a broader audience through \citet{Lin}.
\citet{tollmien_productionofturbulence} analyzed the perturbations in
laminar shear flows as solutions of the Rayleigh equation and
established the dynamic significance of the location where the
averaged flow velocity profile matches the phase speed of the
perturbation, known as critical layer. In the 1940s and 1950s novel
results on flow stability were spreading and igniting interest among
fluid dynamicists \citep{MilesForewordToDrazinReid}. Surveying the
ideas lying around, \citet{miles} postulated that the laminar flow
analysis and results of \citet{tollmien_productionofturbulence} apply
to the generation of surface waves by a wind that is turbulent.
However, \citet{miles}   offered no
physical justification of the proposition and no testable predictions
from it, which greatly influenced the perception, including Miles'
own, of that proposition.  Sir James Lighthill, at the time already an
authority in fluid mechanics and whose life and work are celebrated at
this meeting, wrote favorably of \citet{miles} and referred to it as
'the Miles theory.' Still, he viewed \cite{miles} as a mathematical
construct with somewhat tenuous relation to the physical world and
hence in \cite{lighthill} he sought an interpretation for it. Like
\citet[p.\ 217]{UrsellReview1956}, \citet[p.\ 385]{lighthill}
recognized the agreement between theory and experiment as essential to
gaining insight into the wind waves generation.  With the
observational support still absent, however, \citet[p.\ 1]{miles_wow}
viewed {\em 'Lighthill's opinion [as] too optimistic'} and cautioned
\cite[p.\ 166]{miles5} that Lighthill's \cite{lighthill} endorsement
of the mathematical theory in \cite{miles} has been {\em 'interpreted
  with less reservation than either he or the writer might have
  wished.'}  \citet{MilesAtJHU1998} considered the concept of critical
layer in wind-wave interaction only as a {\em 'convenient mathematical
  notion'.}  Uncertain about the suitability of the ideas in the 1957
paper, Miles' later work proceeded with modifications of that paper's
concept rather than with its experimental verification,
interpretation, potential physical impacts or operational use.
Besides Miles, others have explored such modifications, 
e.g.\ in specific corners of the parameter space, 
the work of \citet{Sajjadi2014} among the latest. 

\citet{miles} postulation that the surface waves occur 
as growing perturbations of the wind profile and are thus 
described by Rayleigh equation, is now superseded by a 
derivation that renders moot the gaps and deficiencies in wind 
input knowledge as perceived by \citet[p.\ 21]{PUSHKAREV201618}. 
Resting solely on the well-confirmed premise  
of small slope waves, \cite{HristovJPO2014} demonstrated 
that the Taylor-Goldstein equation describes the 
wave-coherent component, and thus the wind-wave interaction,
in a stratified turbulent wind. Both the Rayleigh and the 
Taylor-Goldstein equations predict critical layers for a wide 
variety of wind profiles $U(z)$.  The wave-coherent motion there
is associated with a kinetic energy production 
$\langle\tilde{u}\tilde{w}\rangle(\partial U/\partial z)$ 
that is delivered from the mean flow to the waves and is
leading to a wave generation for any of these  profiles.   
Because the logarithmic wind 
profile $U(z) = (u_*/\kappa) \log(z/z_0)$ is sufficiently  
representative of the marine atmospheric boundary layer
\cite{CharnockNature56,Ruggles70,MitsuyasuHondaJFM1982,SolovievKudryavtsev2010}, 
\citet{miles}  offered numerical estimates for the wave growth 
rate for the case of that profile.  While the key concept in 
\cite{miles} has been only illustrated with a logarithmic 
profile, that still invites the misinterpretation, e.g.\ in \citet[p.\ 21]{PUSHKAREV201618},
that the concept in \cite{miles} requires such profile.  It does not. 
Stratification as well as inhomogeneity and non-stationarity  
of the marine atmospheric boundary layer may cause departures     
of the wind profile from its logarithmic shape  \cite{HristovJPO2014}. 
While in such cases the wind-wave interaction still occurs through 
the same physical mechanism of wave-coherent flow with critical layers, 
the wind-to-wave energy 
transfer rates may vary substantially from their values obtained for a
logarithmic profile. Yet, experimental estimates of the wave growth rates
have been compared   with theoretical predictions calculated for the case of 
a logarithmic wind profile.  Discrepancies in such comparisons, although 
not in whole, are related to the unphysical narrowing, as in 
\cite[p.\ 21]{PUSHKAREV201618},   of the theory's scope  to a 
logarithmic wind profile.

Over the decades that followed, review papers have repeatedly
described \cite{miles} as difficult to understand, interpret, validate
or apply
\citep{Bryant1965,StewartRW_PhysFluids1967,BarnettKenyon_Review}, and
as {\it ``...least well understood because of the less-than-intuitive
  nature of the theory''} \citep{CBLAST_Workshop}. With such sentiment
widely shared, the proposition in \cite{miles} has been misconstrued, 
doubted and dismissed.  Regarding the Hibbs' \cite{HibbsWaves} question, that
sentiment may have dissuaded close interest in and may have delayed 
the progress on the wind waves generation problem.

\section{Empirical studies of wind-wave interaction}

With the mechanistic theory of wind-wave interaction deemed
inaccessible, the attention has turned to empirical studies.  There,
the ocean is treated as a rough aerodynamic surface where the waves of
all scales are responsible for the surface roughness and the
variability of the surface's drag. Motivated by the need for
computational efficiency in large scale models of atmospheric dynamics
and ocean circulation, this vast effort has included dozens of
experiments in laboratory, over lakes and over the coastal and the
open ocean and since the 1960s has produced hundreds of papers. It has
been primarily concerned with applied issues, such as establishing
empirical relationships between easy-to-measure environmental
variables, e.g.\ wind speed, vertical scalar gradients, etc., and
variables of ocean-atmosphere exchange, e.g.\ fluxes of momentum,
heat, species, etc. Among the constraints of this empirical approach
is the inability to distinguish between momentum or energy transferred
to the waves and those transferred to the currents, leaving a
significant blind spot for numerical modeling and forecasting.  Among
the main interests pursued within this effort have been the empirical
estimates of the air-sea exchange coefficients. The scatter of these
estimates, however, has not been reduced over time, despite the
improved instrumentation and the accumulation of more statistics
\citep{donelan}. The latter suggests that at the core of the
uncertainty is the complex dynamics of the air flow laying outside of
the empirical framework, rather than experimental imperfections or
isolated statistical aberrations.

A related line of inquiry within the empirical approach on wind-wave
interaction and characterization of the marine atmospheric boundary
layer has been to seek violations of the Monin-Obukhov similarity
theory over the ocean, e.g.\ in the shape of the wind profiles or in
the budgets of momentum flux and kinetic energy, ascribe them to the
influence of the surface waves
\cite{EdsonTKE,EdsonFairallSullivan2006IUTAM,SjoblomSmedmanTKE2002,JanssenTKE,HogstromHristov2013},
and use these violations to understand and quantify the dynamic
wind-waves exchange.  The underlying assumption of such studies is
that no violations of the Monin-Obukhov similarity theory occur over
land, so that any violations of that theory can be interpreted with
certainty as a dynamic signature of the surface waves. Once found and
their patterns discerned, these violations were to be incorporated in
an updated theory, where the universal gradient functions 
$\{\phi_m(z/L), \phi_h(z/L), \ldots\}$ of the Richardson number 
$(z/L)$, where $L$ is the Obukhov mixing length,
are modified to account for
the waves' presence \citep{EdsonFairallSullivan2006IUTAM}.  Such
logic, however, ignores multiple reports of discrepancies between the
said theory and observations over land
\cite{PanofskyAtmBoundaryLayerBelow150Meters,WyngaardAtmosphericTurbulence,Foken50YearsMOIUTAM}.
Furthermore, the assumption that the wave influence would depend
solely on the Richardson number and be independent of any variables
characterizing the sea state, appears unjustified.  The research
strategy just outlined, has detected no wave signature in the study of
\citet{EdsonTKE}. Consistent with the report of
\citet{CharnockNature56} and the studies of
\citet{Ruggles70,MitsuyasuHondaJFM1982,SolovievKudryavtsev2010} and of
\citet{BergstromSmedman}, the \citet{EdsonTKE} experiment determined
that the kinetic energy balance is virtually unaffected by the
presence of waves. Their proposed explanation was that the instruments
have been positioned above the layer affected by the wave influence,
commonly referred to as the wavy boundary layer (WBL).  \citet
{SjoblomSmedmanTKE2002} point out that the research strategy produces
{\em ``seemingly contrasting results''}, that are circumstantial and
inconclusive.

In its broadest formulation the empirical approach would select a
statistical characteristic of the atmospheric boundary layer, be that
dissipation rate, velocity structure function, velocity spectra, etc.,
search for differences in that characteristic in measurements over
land and over waves and from these differences seek insights into
the wind-wave interaction.  \citet{SchacherTKE} have found no detectable
wave signature in the rate of kinetic energy dissipation,
\citet{VanAttaChenStructureFunctionsOverOcean} have failed to identify
such signature in the wind velocity structure function.
\citet{Pond1963turbulence,Pond1966spectra,weiler_burling,StewartRW_PhysFluids1967,SolovievKudryavtsev2010}
have reported no peak at the wave frequency in the wind velocity
spectra.  All these negative, yet informative results indicate that
the mechanical air-sea interaction is both intense and subtle. The
size and energy of the waves generated by a storm indicates the
interaction's intensity.  Yet the interaction does not manifest itself
in quantities commonly used to characterize the atmospheric boundary
layer (wind profiles, spectra, structure functions, kinetic energy
budget, etc.), hence the subtlety.  That subtlety may partially hold
the answer to the question posed by \citet{HibbsWaves}. The negative
results listed above suggest that the selected characteristics 
of the atmospheric boundary layer are insensitive to wave influence 
and that choosing a different set of informative
variables and analysis techniques is essential for gaining a physical
insight into the wind-wave coupling.

\section{Mechanistic perspective and its observational validation}

The generation of ocean waves results from the interaction between two
random fields, the air flow and the compliant water surface. Given the
surface elevation ${\eta}$, the vertical velocity $\dot{\eta}$, and
the atmospheric pressure $p_0$ at a point of the water surface, the
averaged energy exchange rate there is $\langle p_0
\dot{\eta}\rangle$.  The latter suggests that the wave-coherent
pressure, and generally the wave-coherent motion in the air, carries
the interaction.  Therefore, the dynamic equations of that motion
would be the key to formulating the mechanistic theory, while
discerning that motion's dynamic and statistical patterns from field
measurements would be key to verifying the theory.
   
The wave-coherent motion, however, has been perceived as elusive to
both define and detect.  \citet[p.\ 108]{phillips} thought of it: {\em
  ``In physical space, the induced pressure at any point on the
  surface of a random wave field is a rather ill-defined functional of
  both the wind and the wave fields, and it is not easy to separate
  this from the turbulent contribution.''}  \citet[p.\
101]{hasse_dobson} saw experimental challenges: {\em ``This means that
  in the air above the sea, where there is usually a mean wind many
  times stronger than the wave-induced flows, the wave-induced motions
  are hard to detect.''}, as did \citet[p.\ 72]{komen}: {\em ``The
  measurement of the energy transfer from wind to waves is, however, a
  very difficult task, as it involves the determination of the phase
  difference between the wave-induced pressure fluctuation and the
  surface elevation signal...''}


\begin{figure}[t]
 \vspace*{3.7in}
 \includegraphics{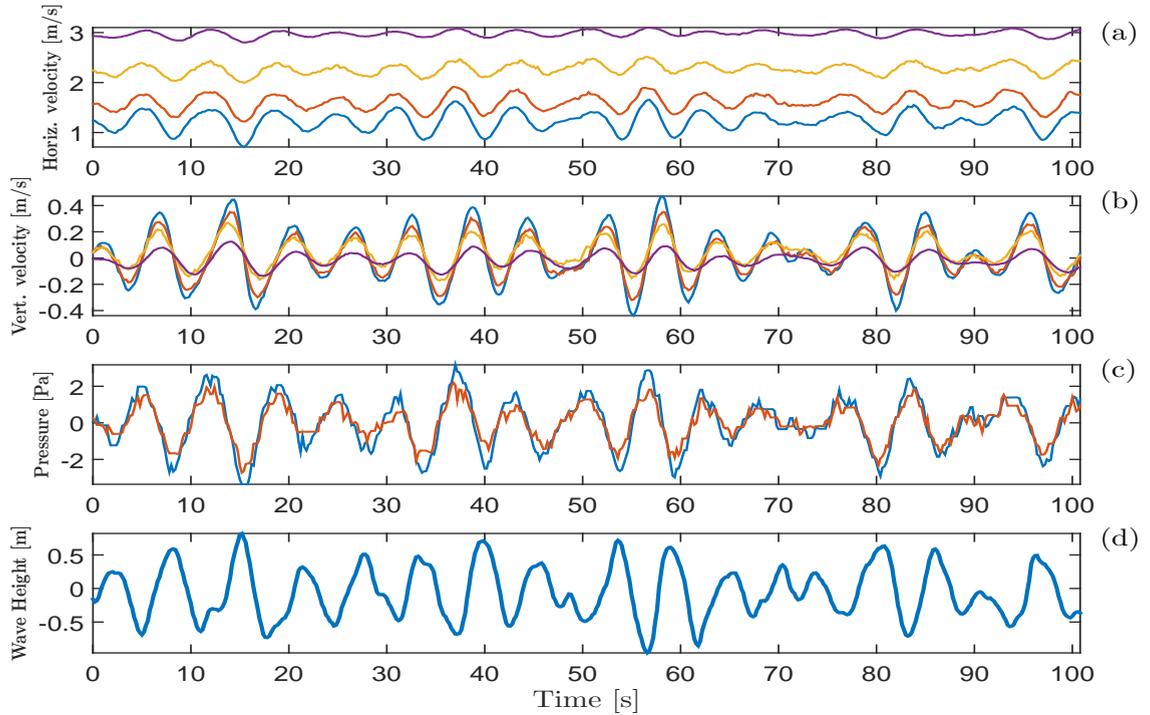}
 \caption{Wave modulation of the wind during Coupled Boundary Layers Air-Sea Transfer 
  (CBLAST) experiment. 
	(a) Horizontal wind velocity measured by 4 ultrasonic anemometers 
	positioned between 5m and 18m from the surface
	in the order blue (the lowest), red, yellow, magenta (the highest).
	(b) Vertical wind velocity, with colors matching the same order of the 
	instruments. (c) Pressure, blue (lower) and red (higher). 
	(d) Surface wave elevation.}
 \label{fig:CblastWC}
 \end{figure}

When the wind speed and the turbulence intensity in the air are
relatively low, the wave-coherent modulation of the wind is directly
observable in unprocessed signals of air velocity and pressure (Figure
\ref{fig:CblastWC}). With the increase of wind speed, turbulence
begins to dominate and the identification of wave-coherent modulation
requires statistical tools
\citep{hristovPRL98,hristovNature,MelvilleCriticalLayer,HristovJPO2014}.

Consider a signal of an air flow variable $u(t)$ that consists of a
wave-correlated component $\tilde{u}(t)$ and a uncorrelated component
$u^\prime(t)$, so that $u = \overline{u} + \tilde{u}(t) +
u^\prime(t)$.  Consider a wave signal $\eta(t)$ as consisting of a
finite-number narrow-band components $\{\eta_n(t)\}$, i.e.\ $\eta(t) =
\sum\limits_n \eta_n(t)$, for which $\langle\eta_k \eta_l\rangle \sim
\delta_{kl}$.  Then, the wave-coherent component $\tilde{u}(t)$ can be
obtained as a projection of the measured signal $u(t)$ onto the vector
space of all wave-coherent signals $W={\rm span} \{\eta_n,
\hat{\eta}_n\}$, where $\hat{\eta}_n$ is the in-quadrature counterpart
of $\eta_n$ obtained via the Hilbert transform, i.e.\
\begin{equation}
\tilde{u}(t) =  \sum_n 
                \left( \frac{\langle u(t) \eta_n(t)\rangle }{|| \eta_n||^2} \eta_n(t)
							       	+
								       \frac{\langle u(t) \hat{\eta}_n(t)\rangle }{|| 
											 \hat{\eta}_n||^2} \hat{\eta}_n(t)
                \right).
\label{wcfilter}
\end{equation}
The filter $\widetilde{(\cdot)}$ defined this way has the properties
$\widetilde{(u)} = \tilde{u}$, $\widetilde{(\tilde{u})} = \tilde{u}$,
$\widetilde{(u^\prime)} = 0$, and $\langle u^\prime
\tilde{u}\rangle=0$.  Such filter both defines and detects
wave-coherent fields in the air flow.  It is suitable for deriving the
dynamic equations for the wave-coherent flow and for retrieving the
time series of the wave-coherent variables in measured signals
\citep{hristovPRL98,hristovNature,HristovJPO2014}.

Referring to \cite{hristovNature}, \cite{HristovJPO2014} and
\cite{Hristov_WileyEncyclopedia} for analytic and experimental
details, below we outline the foundations of the theory there, its
experimental verification, its predictions, and their implications.
Invoking only the assumption of small slope waves,
\cite{HristovJPO2014} employs the filter (\ref{wcfilter}) to arrive to
the dynamic equations for the wave-coherent flow, thus extending the
shear flow instability analysis to the interaction of waves with a
turbulent wind. As the transfer functions of the wave-coherent fields
in the air are invariant with respect to the wave spectrum and can be
produced both theoretically and experimentally, they are well suited
for validating theory from measurements.  Therefore, the wave-coherent
flow structure and dynamics can be described through the transfer
functions between the waves and the air velocity. Matching analytical
and experimental transfer functions would support the theory that
predicted them or, conversely, would explain the experiment that
produced them.  The theory predicts that at low wind speeds 
the wave-coherent flow may form a regime with Stokes drift 
(e.g.\ as in Figure \ref{fig:CblastWC}), causing momentum flux 
from waves to wind, and a regime with critical layers at moderate 
and high wind speeds 
\cite{hristovNature,MelvilleCriticalLayer,HristovJPO2014}, 
causing momentum and energy fluxes from wind to waves. A distinct 
flow feature is the discontinuity of the velocity's phase at the 
critical height \cite{hristovNature,Hristov_WileyEncyclopedia}.

Since coastal ocean and laboratory are much more accessible than the
open ocean for wind-wave interaction studies, it is pertinent to
recognize the constraints of such environments for observing the critical
layer signature.  The dispersion of waves in finite-depth water
$\omega^2 = gk \tanh(kd)$, where $g$ is the acceleration of gravity
and $d$ is the water depth, establishes a maximum for the waves' phase
speed $c_{max} = (gd)^{1/2}$.  When waves transition from deep to
shallow water, the refraction may misalign the wind and the waves, 
requiring that the relative direction is included in estimating the 
wave age. Furthermore, for the range of wave frequencies $[0, \infty]$, the
range of possible phase speeds contracts from $[0, \infty]$ to $[0,
\sqrt{gd}]$. For a frequency range $\Delta \omega$ the corresponding
phase speed range in deep water $\Delta c_{\rm deep} $ contracts to a
phase speed range in shallow water $\Delta c_{\rm shallow} $,   so that 
$\Delta c_{\rm shallow} < \Delta c_{\rm deep}$.  In turn, the range of
critical heights $\Delta z_{c,{\rm deep}}$ corresponding to the same
frequency range $\Delta \omega$, contracts as well to $\Delta
z_{c,{\rm shallow}}$, i.e.\ $\Delta z_{c,{\rm shallow}} < \Delta
z_{c,{\rm deep}}$, as does the range of wave ages $\Delta(c/u_*)$.  
Compressing the critical layer features of a range
of wave modes $\Delta \omega$ into a thinner  critical heights range
and narrower wave age intervals may cause these features to become 
poorly resolvable, while  the increased wave steepness in shallow water 
may enhance nonlinearities and distort any critical layer 
signature.

Measurements over deep water waves, conducted during the Marine 
Boundary Layer Experiment \cite{hristovNature,Hristov_WileyEncyclopedia} 
and the High Resolution Air-Sea Interaction Experiment 
\cite{MelvilleCriticalLayer,HristovJPO2014}, along with the 
analytic results  in \cite{HristovJPO2014,Hristov_WileyEncyclopedia}
 established that:  
\begin{enumerate}[(i)]
\item 
  The wind-wave interaction does occur through a wave-coherent flow in
  the air and the critical layer pattern of phase discontinuity in
  that flow is sustained, thus identifying the mechanism responsible
  for wind wave generation,
  \cite{hristovNature,MelvilleCriticalLayer,Hristov_WileyEncyclopedia}.

\item 
  A Stokes drift regime is observed at low wind conditions, associated
  with weak, yet pronounced wave-to-wind momentum flux
  \cite{HristovJPO2014}.
 
\item 
  Relying only on the assumption of small slope waves, that is,
  $k_p\sigma_\eta \ll 1$, where $k_p$ is the wave number of the
  spectral peak and $\sigma_\eta = \langle\eta^2\rangle^{1/2}$ is the
  variance of the sea surface, the analysis in \cite{HristovJPO2014}
  shows that the budgets of second order moments, e.g.\ kinetic energy
  and momentum, apply separately to the wave-correlated and
  wave-uncorrelated motions in the wind.  As the phenomenology of the
  atmospheric boundary layer's kinetic energy budget is the essence of
  the popular Monin-Obukhov similarity theory (MOST), the separated 
	budgets show how the waves modify the predictions from MOST.  

\item 
  The key measure of the waves' dynamic influence  on the air
  flow is the ratio of the production terms in the kinetic energy
  budgets for the correlated $\langle\tilde{u}\tilde{w}\rangle(\partial U/\partial z)$
	and uncorrelated $\langle u^\prime w^\prime\rangle(\partial U/\partial z)$
	motions, which in turn
  is expressed as a ratio of the wave-supported
  $-\langle\tilde{u}\tilde{w}\rangle$ and total $-\langle u w\rangle =
  u_*^2$ momentum fluxes, i.e.\ $-\langle\tilde{u}\tilde{w}\rangle /
  u_*^2$, \cite{HristovJPO2014}.

\item 
  At heights available for atmospheric measurements $z \geq 10^4 z_0$,
  where $z_0$ is the aerodynamic roughness length of the sea surface,
  typically between $2\times 10^{-4} {\rm m}$ and $5\times 10^{-4}
  {\rm m}$, both theory and experiment find that the ratio
  $-\langle\tilde{u}\tilde{w}\rangle / u_*^2$ is small, of the order
  of 5\%, \cite{HristovJPO2014}.

\item
  Consequently, the wave contribution to the sea surface drag coefficient
  $C_D$, the wave-induced modification of the kinetic energy budget,
  the apparent wave-enhanced imbalance between production and
  dissipation, the wave contribution to the departures from the
  predictions of the Monin-Obukhov similarity theory and the
  wave-induced bending of the wind profile are also small and thus
  virtually undetectable next to other physical influences modifying
  these characteristics, in agreement with empirical studies,
  \cite{HristovJPO2014}.

\item 
  The explicit forms of the wave-supported momentum and kinetic energy
  fluxes indicate that a wave frequency spectrum $\omega^{-\beta}$,
  through relaxation, converges to a spectral slope $4 \leq \beta \leq
  5$, \cite{HristovJPO2014}.
 
\end{enumerate}

These findings explain the negative results in the vast body of
empirical studies that sought a wave signature in standard
characteristics of the flow over waves, such as a wind profile, a
dissipation rate, a kinetic energy balance, a momentum flux, or a
variation of the drag coefficient with sea state.  They show that the
predictions of the Monin-Obukhov similarity theory are virtually
insensitive to the waves, letting the conclusion that the theory
adequately describes the marine atmospheric boundary layer, yet it is a poor
instrument for detecting and studying wind-wave coupling.  The
demonstrated agreement between theory and experiment unifies the
mechanistic and empirical perspectives on wind-wave interaction.

\section{Experimental and numerical studies of  wave growth rate and wave energy evolution}

Wave growth rate and wave energy evolution under wind forcing are of
primary interest for  wave modeling and forecasting \cite{PUSHKAREV201618}. 
The wave growth rate, rather
than the air flow pattern, has been central to the thinking and
studies of wind waves to the extent that a discrepancy with observed
growth rates sufficed for \citet[p.\ 228]{Donelan1990} to dismiss the
available theory: {\em "Thus, our knowledge of the wind input to a
  spectrum of waves is still rather primitive. The theoretical ideas
  of the fifties have not been capable of explaining the observed
  growth rates and no essentially new ideas have followed."}  Yet,
experimental and numerical estimates of growth rates and of energy
evolution depend on the constraints and the uncertainties in the
methods used to obtain them. Below, these constraints and
uncertainties are outlined for proper interpretation of wave growth
rate estimates' relation to a particular theory.

As a wave generation through random force \citep{phillips57} or
through shear flow instability
\citep{tollmien_productionofturbulence,miles}, would lead to a
different growth rate and a different evolution of the wave field's
energy, experiments have sought to identify wind-wave generation
scenarios through growth rate or energy evolution measurements.
\citet{davis70} noted, however, that the growth rate is a function of
the pressure distribution on the surface, yet since the pressure
distribution does not uniquely specify the flow structure in the air,
it does not determine the mechanism of wind-wave interaction.  As for
the evolution of the wave field energy observed in an experiment, it
depends not only on the wind-wave interaction mechanism, but also on
the history and spatial distribution of the variable wind forcing as
well as on the action of non-linearity that redistributes the wave
energy across the spectrum. Because different wave modes draw momentum
and energy from the wind at different rates, a wave field developing
with non-linearity may have evolution of its energy significantly
different from a wave field developing under the wind forcing alone,
i.e.\ when each mode retains exactly the energy received directly from
the wind.  Consequently, like the wave growth rate, the observable
wave field energy evolution lacks the certainty necessary for
identifying the physics of wind-wave interaction.


\begin{figure}[t]
 \vspace*{5.4cm}
 \includegraphics{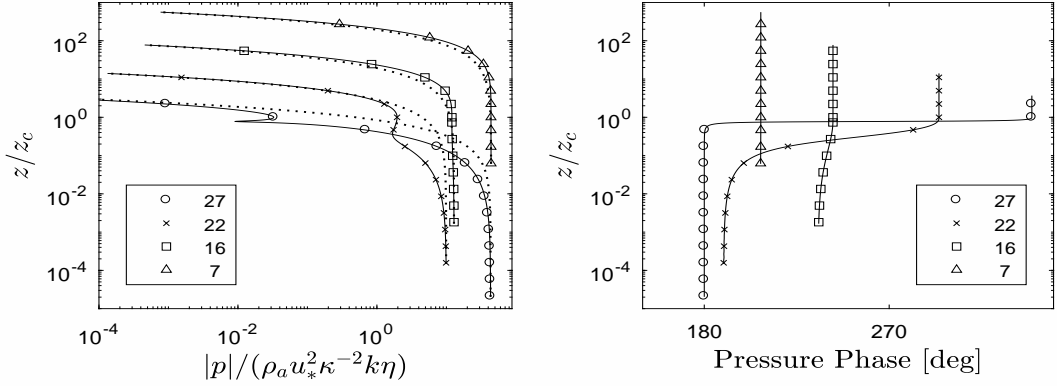}
 \caption{The normalized  magnitude 
          $|p| / (\rho_a u_*^2 \kappa^{-2} k\eta)$ (left), 
					and phase (right) of the wave-coherent pressure vs.\ the 
					distance from the surface normalized by the critical 
          height $z/z_c$, obtained from a solution of the Rayleigh 
					equation \cite{hristovNature,HristovJPO2014,Hristov_WileyEncyclopedia}.
					The symbols $\circ,\times,\square,\triangle$ distinguish 
					different values of the wave age $c/u_*$. 
					The dotted lines show exponents that match the pressure magnitude 
					at the endpoints and exhibit a notable departure in the middle.}
 \label{fig:NumSolPressure}
 \end{figure}

 Multiple experiments
 \cite{elliott72,snyder74,SnyderDobsonElliottLong,HasselmannBosenberg,wilczak_pressure,donelan_wow,donelan_pressure,SavelyevHausDonelan_pressure}
 have been carried out to measure the wind-to wave energy transfer
 rate as a correlation between pressure and surface velocity, i.e.
 $\langle p_0 \dot{\eta}\rangle$ and compare them with predictions
 from theory. Pressure measurements on the surface risk wetting the
 sensor or distorting the readings with protective film, etc.
 Commonly, pressure measurements are conducted at some finite height
 and later extrapolated to the surface, assuming exponential decay of
 the pressure's magnitude scaled by the wave number $k$, i.e.\
 $\propto e^{-kz}$, and no change in phase with height, as in
 \cite[p.\ 443]{elliott72}, \cite[p.\ 508]{snyder74}, \cite[p.\
 24]{SnyderDobsonElliottLong}, \cite[p.\ 397, p.\ 405, p.\
 407-409]{HasselmannBosenberg}, \cite[p.\ 1020]{wilczak_pressure},
 \cite[p.\ 190]{donelan_wow}, \cite[p.\ 1174]{donelan_pressure},
 \cite[p.\ 1334]{SavelyevHausDonelan_pressure}.  However, a theory of
 wind-wave interaction through wave-coherent flow predicts the
 wave-coherent pressure from solutions of the Taylor-Goldstein equation
 \cite{hristovNature,HristovJPO2014,Hristov_WileyEncyclopedia}, which
 depends on the dimensionless aerodynamic surface roughness
 $\Omega=y_0 g \kappa^2/u_*^2$ and the wave age parameter $c/u_*$
 rather than on the wave number $k$, and exhibits a dependence on
 height that is distinctly different from exponential (Figure
 \ref{fig:NumSolPressure}).  The dotted lines in Figure
 \ref{fig:NumSolPressure} show exponents that match the pressure
 predicted from the Rayleigh equation at the end points and notably
 depart, by up to a factor of 10, from it in the middle. Furthermore,
 the assumption of phase independent of height does not hold for a
 range of wave ages. Consequently, an assumption of wavenumber-scaled
 exponential pressure decay with constant phase is inconsistent with
 the theory that it is employed to test and distorts the pressure
 extrapolation to the surface. Along with the random variability of
 the surface roughness $\Omega$, the distortive pressure extrapolation
 contributes to observed discrepancies between theoretical and
 experimental estimates of the wind-to-wave energy transfer rates.
 Furthermore, the wind-to-wave energy transfer depends on the wind
 profile shape $U(z)$ through the factor $U^{\prime\prime}/U^{\prime}$
 \cite{tollmien_productionofturbulence,miles2,HristovJPO2014}.  The
 wind's non-stationarity and spatial inhomogeneity along with atmospheric 
stratification deform the wind
 profile \cite[p.\ 3192]{HristovJPO2014} and through the factor
 $U^{\prime\prime}/U^{\prime}$ add to the variability of the
 wave-supported energy flux. Although the available theory explains
 such variability, that variability has been unaccounted in any of the
 field experiments listed above.

 Since the 1970s, RANS and LES models for the flow over waves have
 been used evaluate wind-to-wave energy transfer.  Such models rely on
 concepts (eddy viscosity, diffusivity) and tools (closures and
 sub-grid parametrization schemes) developed for describing turbulent
 flows.  The strong sensitivity of modeling results to the choice of
 closure \cite{Al-zanaidi_Hui_1984} and the fact that no second-order
 closure model has detected any critical layer flow features
 \cite{mastenbroekThesis,SullivanMcWilliamsMoeng}, has led
 some to question the adequacy of these closures.  \citet[p.\
 117]{phillips} observed that {\em "Closure schemes in turbulent shear
   flow are still rather ad hoc and different methods, which may be
   reasonably satisfactory in other flows, give very different results
   when applied to this problem.  The situation is not one in which
   firmly established methods lead to results that one might seek,
   with some confidence, to verify experimentally. On the contrary,
   because of sensitivity of results to the assumptions made, the air
   flow over waves appears to provide an ideal context to test the
   theories of turbulent stress generation themselves."} As outlined
 below, the inadequacies in such models extend beyond the deficiencies
 in closures and sub-grid parametrizations.

 Starting from \citet{Al-zanaidi_Hui_1984}, a number of studies have
 modeled the flow over a Stokes wave, that is
\begin{equation}
\eta(x) = -a \cos(kx) - (ka^2/2)\sin^2(kx),
\nonumber
\end{equation}
instead of over a monochromatic wave $\eta = ae^{-{\rm i} (kx - \omega
  t)}$.  Such studies ascribe to a single mode $k$ the energy flux to
two wave modes, $k$ and $2k$, thus producing a spurious enhancement of
the wave growth rate.

Contrasting with the  widely held views about turbulence, 
the wave-induced flow is anisotropic at both large
and small scales.  Its directions of distinct significance are the
direction of wave propagation, and the vertical direction, along which
the wave signature decays. The critical layers, which both the theory
and experiment show to be dynamically essential in wind-wave
interaction, create a fine structure in the wind. The wave-induced
flow may experience a large variation over a short distance near the
distinctly anisotropic critical layer \cite[Figure 1e]{hristovNature}.
The critical layers of different wave modes are densely stacked on top
of each other along the vertical coordinate.  Closures and sub-grid
parameterizations have been proposed with regard to the assumed
properties of turbulence and none with regard to the properties of the
wave-induced flow, which differs from turbulence in its scales and 
symmetries.  Resolving the wave-induced flow with
multiple critical layers \cite[Figure 1e]{hristovNature} requires a
fine spatial grid.  The 40 cells along the vertical coordinate used by
\citet[p.\ 234]{Al-zanaidi_Hui_1984} or LES grids with $\Delta z \geq
1m$ near the surface, \cite{HristovJPO2014}, define a domain discretization  
 too coarse to capture
the wave-induced flow structure just described.  Although such coarse grids 
may resolve the large scale motion of the atmospheric
Stokes drift and the associated wave-to-atmosphere momentum transfer,
they suppress the critical layers and the concomitant wind-to-wave
transfer. This way the grid size selects the elements of wind-wave
dynamics that are retained and that are ignored in a LES model.

\section{Conclusion}

\citet{HibbsWaves} contemplated the causes that kept the wind waves 
generation mechanism as a longstanding open physical problem. 
Within the century between the pioneering work of
\citet{kelvin} and \citet[p.\ 228]{Donelan1990} dismissal of the contemporary
theory for its inability to explain the experimental growth rates,
theoretical, empirical and numerical approaches have been employed
inconclusively.  Over the last one and a half decade, analytic,
numerical and open ocean experimental results
\cite{hristovNature,MelvilleCriticalLayer,HristovJPO2014,Hristov_WileyEncyclopedia}
have confirmed that the mechanism of wind-wave interaction through
wave-coherent flow is indeed active and the critical layer pattern in
that flow is persistent and pronounced.  Such findings offer a new
mechanistic perspective on the constraints and challenges through the
evolution of ideas and methods relied on to study the problem.  Among
these have been:
 \begin{enumerate}[(i)]
\item 
  The lack of physical justification and testable predictions, as well as the 
  {\em 'less-than-intuitive'} \cite{CBLAST_Workshop} analytic details,
  have made the theoretical ideas to be misconstrued, doubted and dismissed,
  \citep{Bryant1965,StewartRW_PhysFluids1967,miles5,BarnettKenyon_Review,Donelan1990,MilesAtJHU1998,miles_wow}.

\item 
  Instead of seeking air flow patterns that uniquely do identify the
  wind-wave interaction mechanism, the experimental studies
  \cite{elliott72,snyder74,SnyderDobsonElliottLong,HasselmannBosenberg,wilczak_pressure,donelan_wow,donelan_pressure,SavelyevHausDonelan_pressure}
  have been focused on wave growth rates, that do not \cite{davis70}.

\item 
  Improper choice of the wave number $k$ instead of the wave age
  $c/u_*$ as a governing parameter, distortive exponential
  extrapolation to the surface, variation of the wind-to-wave energy
  flux through the factor $U^{\prime\prime}/U^{\prime}$ due to
  deviation of the wind profile from its logarithmic shape, and
  possibly other experimental imperfections and uncertainties, are
  contributors to the discrepancy \cite{Donelan1990} between
  theoretical and measured wave growth rates.

\item Empirical studies have been concerned with air flow
  characteristics weakly affected by the surface waves, e.g. wind
  profiles, kinetic energy balance, structure functions and spectra,
  etc., thus pursuing a wave signature that is virtually undetectable in
  experiments \cite{EdsonTKE,HristovJPO2014}.

\item The wave-coherent flow (Figure \ref{fig:CblastWC}), a key to the
  mechanistic description of wind-wave coupling, has been perceived as
  elusive to both define and detect in a field experiment \cite[p.\
  108]{phillips}, \cite[p.\ 101]{hasse_dobson}, \cite[p.\ 72]{komen}.

\item Numerical studies have relied on concepts, tools and
  computational grids suitable for turbulent flows. As turbulence
  differs from the wave-coherent flow in its scales and symmetries,
  resolving the wave-coherent flow, and by extension the wind-wave
  interaction, requires computational grids much finer than those used
  so far in LES   \cite{HristovJPO2014}.

\end{enumerate}
These constraints and challenges may offer at least a partial answer
to the question posed by Hibbs' \cite{HibbsWaves}, that is,   
why for more than a century understanding the dynamics of wind-wave interaction has 
been a tenacious physical problem.

\bibliographystyle{elsarticle-harv}

\bibliography{PIUTAM_WW_2017_HristovReferences}

\end{document}